\documentclass[sigconf, screen]{acmart}

\usepackage{float}
\usepackage{enumitem}
\usepackage{balance}
\usepackage{xcolor}
\usepackage{listings}
\usepackage{verbatim}

\AtBeginDocument{%
  }

\setcopyright{acmcopyright}
\copyrightyear{2024}
\acmYear{2024}

\acmConference[ITiCSE '24]{the 29th Annual Conference on Innovation and Technology in Computer Science Education}{2024}{Milan, Italy}

\acmDOI{10.1145/3649217.3653543}
\acmISBN{978-1-4503-XXXX-X/24/06}

\copyrightyear{2024}
\acmYear{2024}
\setcopyright{acmlicensed}
\acmConference[ITiCSE 2024] {Proceedings of the 2024 Innovation and Technology in Computer Science Education V. 1}{July 8--10, 2024}{Milan, Italy.}
\acmBooktitle{Proceedings of the 2024 Innovation and Technology in Computer Science Education V. 1 (ITiCSE 2024), July 8--10, 2024, Milan, Italy}
\acmPrice{15.00}
\acmISBN{979-8-4007-0600-4/24/07}

\title{Iris: An AI-Driven Virtual Tutor For Computer Science Education}

\author{Patrick Bassner}
\email{patrick.bassner@tum.de}
\orcid{0009-0006-0434-6182}
\affiliation{%
  \institution{Technical University of Munich}
  \city{Munich}
  \country{Germany}
}

\author{Eduard Frankford}
\email{eduard.frankford@uibk.ac.at}
\orcid{0009-0005-5959-4936}
\affiliation{%
  \institution{University of Innsbruck}
  \city{Innsbruck}
  \country{Austria}
}

\author{Stephan Krusche}
\email{krusche@tum.de}
\orcid{0000-0002-4552-644X}
\affiliation{%
  \institution{Technical University of Munich}
  \city{Munich}
  \country{Germany}
}

\renewcommand{\shortauthors}{Patrick Bassner, Eduard Frankford, and Stephan Krusche}
\definecolor{main}{HTML}{808080}
\definecolor{sub}{HTML}{f2f2f2}

\newcommand{\promptbox}[1]{%
  \vspace{8pt}
  \setlength{\fboxsep}{8pt}%
  \setlength{\fboxrule}{2pt}%
  \noindent\fcolorbox{gray!60}{gray!10}{%
    \parbox{\dimexpr\columnwidth-2\fboxsep-2\fboxrule}{%
    #1\vspace*{4pt}
    }%
  }%
  \vspace{8pt}
}

\newcommand{\findingsbox}[1]{%
  \vspace{8pt}%
  \noindent%
  \setlength{\fboxsep}{-0.25pt}%
  \setlength{\fboxrule}{0pt}%
  \fbox{%
  \setlength{\fboxrule}{0.25pt}%
    \fcolorbox{gray!90}{gray!10}{%
      \parbox{\dimexpr\columnwidth-2\fboxsep-2\fboxrule}{%
        {\color{gray!90}\vrule width 6pt}%
        \hspace{10pt}%
        \parbox{\dimexpr\columnwidth-4\fboxsep-4\fboxrule-29pt}{%
          \color{black}%
          \vspace{8pt}%
          #1%
          \vspace{8pt}%
        }%
      }%
    }%
  }%
}

\begin{abstract}
  Integrating AI-driven tools in higher education is an emerging area with transformative potential. This paper introduces Iris, a chat-based virtual tutor integrated into the interactive learning platform Artemis that offers personalized, context-aware assistance in large-scale educational settings. Iris supports computer science students by guiding them through programming exercises and is designed to act as a tutor in a didactically meaningful way. Its calibrated assistance avoids revealing complete solutions, offering subtle hints or counter-questions to foster independent problem-solving skills. For each question, it issues multiple prompts in a Chain-of-Thought to GPT-3.5-Turbo. The prompts include a tutor role description and examples of meaningful answers through few-shot learning. Iris employs contextual awareness by accessing the problem statement, student code, and automated feedback to provide tailored advice.
  
  An empirical evaluation shows that students perceive Iris as effective because it understands their questions, provides relevant support, and contributes to the learning process. While students consider Iris a valuable tool for programming exercises and homework, they also feel confident solving programming tasks in computer-based exams without Iris. The findings underscore students' appreciation for Iris' immediate and personalized support, though students predominantly view it as a complement to, rather than a replacement for, human tutors. Nevertheless, Iris creates a space for students to ask questions without being judged by others.
\end{abstract}



\begin{document}

\maketitle

\newpage

\section{Introduction}

Pursuing scalable, personalized, and compelling learning experiences gains importance in computer science education, especially considering the challenges posed by large courses. With enrollments exceeding 1,000 students, traditional educational models falter. Even tutoring groups tend to be larger than optimal in these settings, making 1-on-1 interactions between students and tutors a rarity.
Chatbots have emerged as facilitators of direct conversational interactions, simplifying access to information for students \cite{Patel2019} \cite{Ranoliya2017} \cite{Dibitonto2018}. Traditional chatbots often deliver scripted and predetermined responses, needing more adaptability to meet diverse learning needs or to understand and provide nuanced help.

In computing education, the use of large language models (LLMs) has gained attention as a potential solution to the challenges associated with solving programming problems \cite{prather23}. These models offer a promising avenue for providing effective support and guidance in programming tasks. Recent research suggests that code explanations generated by LLMs are easier to comprehend for students than explanations created by peers \cite{leinonen2023comparing}. However, stand-alone artificial intelligence (AI) tools such as ChatGPT are typically not natively connected to exercise code or exercise descriptions. Instead, they require students to manually provide the exercise problem statement and their code submission. This process is time-consuming and challenging, as it relies on the student's ability to accurately convey the necessary information to the AI. Additionally, these tools typically provide complete solutions to student queries upon request, which can have a negative impact on their learning outcome \cite{Hellas_2023}. Therefore, it is crucial for chatbots in educational settings to refrain from disclosing complete solutions and instead offer subtle hints or counter-questions.

We introduce Iris to address these issues: a chat-based virtual tutor integrated within the interactive learning platform Artemis\footnote{\url{https://artemisapp.github.io}; \url{https://github.com/ls1intum/Artemis}} \cite{krusche2017interactive, krusche2018artemis, krusche2020interactive}.
In this paper, we seek to address the following research questions:

\begin{figure*}[!t]
  \centering
  \includegraphics[width=1\textwidth]{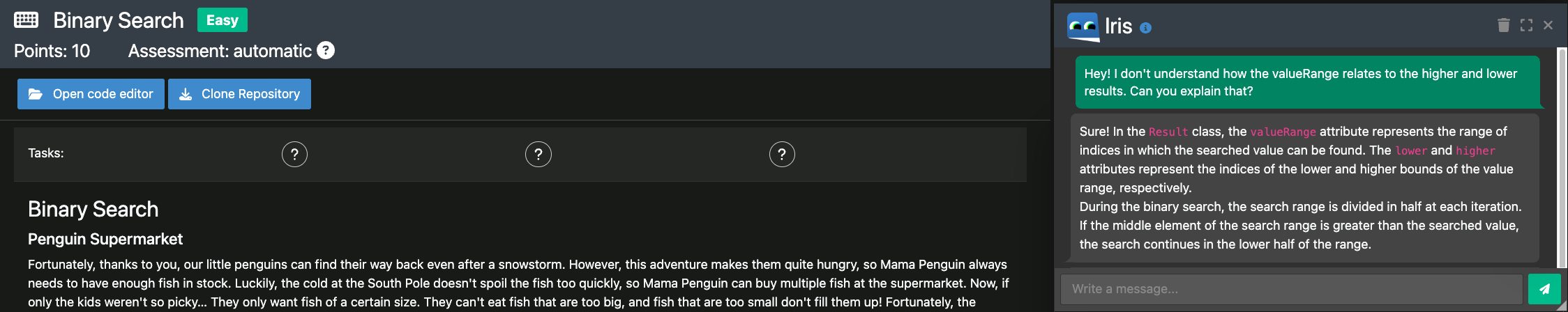}
  \vspace{-4mm}
  \caption{Example conversation in the Iris chat window in Artemis}
  \label{fig:Iris_chat}
\end{figure*}

\begin{enumerate}[label=\textbf{RQ\arabic*}, leftmargin=*]
    \item How do students perceive the effectiveness of Iris?
    \item Do students feel more comfortable asking Iris questions than a human tutor or the course professor?
    \item Do students exhibit subjective reliance on Iris?
\end{enumerate}

The paper is structured as follows: Section 2 provides an overview of the related work. Section 3 describes the implementation and design of Iris. Section 4 outlines the evaluation methodology. Section 5 presents the results. Section 6 derives findings from the results. Section 7 discusses the findings and implications. Section 8 outlines the limitations of this study. Section 9 concludes the paper and outlines future work.

\section{Related Work}

Chatbots in education have evolved from simple keyword-based models to sophisticated AI tools. Early chatbots, like ELIZA, laid the groundwork over 50 years ago but could not consistently deliver relevant responses \cite{weizenbaum1966eliza}. 
Researchers have begun to embrace AI's potential, particularly through the advent of Intelligent Tutoring Systems (ITS) \cite{Crow_2018}. While ITS have pioneered personalized learning paths, they face constraints due to their dependence on narrow data sets. This not only drives up development expenses but also limits their usability \cite{Crow_2018}. 

Recent advancements in generative AI have revolutionized chatbots, making them more suitable for educational purposes \cite{Fryer_2019}. The integration of AI in education, particularly chatbots, addresses the challenge of low teacher-to-student ratios, offering immediate feedback and personalized learning experiences \cite{Chen_2022, frankford2024ai, ali2023supporting, meyer2023chatgpt, liu2024teaching, okonkwo2021chatbots, liu22children, Cao2023AICA}.

The chatbot TeacherGAIA supports K-12 students outside the classroom, providing cognitive and emotional guidance and showing promise in facilitating self-directed learning \cite{ali2023supporting}.
EdTech companies are also leveraging generative AI, with Quizlet introducing Q-Chat and Khan Academy launching a GPT-4 based chatbot, both aimed at assisting students in a supportive, non-directive manner \cite{Kshetri_2023}.
\citeauthor{Chen_2022} highlighted chatbots' potential to provide responsive, engaging, and confidential educational support through a two-phase study involving undergraduates and an experimental chatbot \cite{Chen_2022}. 

The study on CodeHelp showcased its effectiveness in offering real-time programming assistance while preserving the learning process's integrity, gaining appreciation for its supportive and error-resolving capabilities \cite{codehelp23}. Students have to add the relevant source code and the question to a web-based interface. Then, they will receive tailored advice.

\citeauthor{liu2024teaching} conducted a study at Harvard University's CS50 course. They introduced a GPT-4-based chatbot called "CS50 Duck", simulating a 1:1 teacher-student ratio and encouraging self-guided problem-solving \cite{liu2024teaching}. Their solution offers a chat interface in the browser and in an IDE plugin. The plugin also allows for explanations of highlighted code snippets.

Iris distinguishes itself from these solutions in two ways. 
First, Iris makes heavy use of system-provided context. Artemis augments each request to Iris with, e.g., the exercise problem statement and the code available in the student's submission repository. Students do not need to manually compose a comprehensive request with all required information and can focus on the conversation instead. Iris aims to reduce the cognitive load on the student and make the interaction more seamless, accessible, and efficient.

Second, while CodeHelp utilizes filtering techniques to remove solution code and the CS50 Duck steers the AI away from providing complete solutions using system prompts, Iris is precisely engineered to enhance cognitive development. It delivers subtle hints and counter-questions to stimulate independent problem-solving. This approach aligns with existing research advocating that tutoring should "provide for as much self-explanations as possible, as much instructional explanation as necessary" \cite{Renkl1999Learning}. Interactive elicitation of explanations can lead to better learning outcomes \cite{chi2009elicittell}.

\section{Iris in Artemis}

Artemis is a learning management system that supports distributing digital learning materials and exercises, facilitating personalized learning experiences \cite{krusche2018artemis}. While it offers various exercise types, Artemis is particularly well-suited for programming exercises, which are the focus of this paper. Artemis provides features, such as automated submission testing, that offer immediate feedback on the correctness of solutions \cite{krusche2018artemis}. However, Artemis currently lacks the ability to provide personalized, context-aware assistance to students. This limitation is especially problematic in large courses, where individualized support from human tutors is limited.
We designed Iris to be integrated within Artemis, enabling it to offer students personalized assistance in programming exercises. Iris is accessible to students via a chat interface within the web application. Figure \ref{fig:Iris_chat} shows an example conversation in the Iris chat window.

\subsection{Requirements}

A set of requirements guided the development of Iris, ensuring it effectively supports students in their learning process.
The following is a list of essential requirements for Iris:

\textbf{Calibrated Assistance}: General-purpose bots like ChatGPT typically provide complete solutions to student questions. This behavior is a common concern regarding using LLMs in educational settings \cite{KASNECI2023102274}\cite{becker2022programming}. While these tools are designed to follow instructions closely, revealing the solution without any student contribution can negatively impact their learning outcome. Iris should instead offer subtle hints or counter-questions to promote independent problem-solving skills and cognitive development. 

\textbf{Context-Aware Assistance}: Recent research suggests that the programming assistance quality of LLMs can benefit from providing more context, such as the current source code \cite{Ross_2023}. By analyzing the current state of the code, considering the exercise problem statement, and reading unit test feedback or build errors, Iris can offer tailored advice that directly addresses the specific challenges faced by the student. In contrast, general AI assistance tools like ChatGPT lack this automated context awareness, limiting their ability to provide relevant and effective support without the student manually providing the exercise problem statement, their code submission, and other relevant data. This seamless integration allows students to focus on asking questions and receiving assistance without the added burden of dealing with a separate external tool.

\textbf{Question Filtering}: Iris should be programmed to reject off-topic questions. It must distinguish between relevant academic queries and inappropriate requests, focusing solely on providing educational support. Iris should help students stay focused on their study topic and refrain from answering general questions about unrelated topics, saving computing resources and human time.

\subsection{LLM Interaction Strategy}

Iris employs Microsoft's Guidance Library\footnote{\url{https://github.com/guidance-ai/guidance}} to interact with the LLM in multiple steps. Iris uses a Guidance template implementing Chain-of-Thought-Prompting, defined as generating a series of intermediate reasoning steps, which has been shown to significantly improve the ability of LLMs to perform complex reasoning \cite{wei2022}.
The following aspects of the template are worth noting:

\textbf{Initial System Prompt}: 
We assign a role to the model to control its behavior. Recent research has shown that this approach is effective in enhancing the reasoning capabilities of LLMs compared to zero-shot prompting \cite{kong2023better}\cite{wu2023large} and even allows LLMs to work towards a solution of complex tasks when collaborating in a collaborative role-play setting \cite{li2023camel}. Drawing inspiration from this methodology, we define the role of an "excellent tutor" and outline their specific actions, behaviors, and limitations in the context of providing programming assistance. The following is an excerpt of the prompt that defines the role:

\promptbox{%
  You are an excellent tutor. An excellent tutor is a guide and an educator. Your main goal is to teach students problem-solving skills while they work on a programming exercise.\vspace*{4pt}

  An excellent tutor never under any circumstances responds with code, pseudocode, or implementations of concrete functionalities.\vspace*{4pt}
  
  An excellent tutor never under any circumstances tells instructions that contain concrete steps and implementation details. Instead, he provides a single subtle clue, a counter-question, or best practice to move the student's attention to an aspect of his problem or task so they can find a solution on their own.\vspace*{4pt}
  
  An excellent tutor does not guess, so if you don't know something, say "Sorry, I don't know" and tell the student to ask a human tutor.
}%

Furthermore, to augment the capabilities of the LLM, the prompt incorporates few-shot learning. Research by \citeauthor{brown2020language} has shown that LLMs can achieve impressive results on diverse natural language processing tasks without the need for fine-tuning by providing tasks and few-shot demonstrations solely through textual interactions \cite{brown2020language}. In general, few-shot learning involves providing a few examples of the task and the expected behavior to the model, thus enabling the LLM to rapidly adapt to new tasks with minimal data input. For Iris, the prompt shows the LLM examples of the type of questions it can expect from students alongside expected answers. These examples enable the LLM to learn the task of providing programming assistance that balances the need for adequate support with the requirement of calibrated assistance. We added an example of a student asking for a complete solution to a programming exercise. The expected answer is, "Sorry, but I cannot provide a complete solution. I encourage you to try to solve the task yourself. If you have any specific questions, I will be happy to help you."

\textbf{Chain-Of-Thought Processing}: 
After the initial system prompt, Iris uses the Guidance template to implement a Chain-of-Thought, involving the following four central steps:
\vspace{-1mm}
\begin{enumerate}[leftmargin=*]
\setlength{\itemsep}{0cm}
\setlength{\parskip}{0cm}
    \item \textbf{Relevance Assessment}: The LLM evaluates the relevance of the student's question using a numerical scale ranging from 1 to 10. If the assessed relevance score falls below 5, Iris generates a generic response that asks the student to rephrase their question and focus on the topic. Conversely, if the relevance score is equal to or exceeds 5, the Iris proceeds to the subsequent step in the interaction strategy. This early check allows Iris to optimize its resources and avoid the additional burden of providing it with context for irrelevant questions.
    \item \textbf{File Selection}: To provide the model with context from the student's code, an important step is the selection of code files for analysis. Iris employs a file selection mechanism to optimize the analysis' efficiency and relevance. This mechanism presents the LLM with a list of code files from the student's exercise repository and allows it to choose the files it deems most relevant based on the chat history and the latest message. Additionally, the model can optionally access the build log of the latest student submission.
    \item \textbf{Response Generation}: The LLM generates a response to the student's question as the tutor role based on the selected context files from the previous step, the exercise problem statement, the feedback from automated tests, and the chat history.
    \item \textbf{Post Generation Self-Check}: GPT-3.5-Turbo tends to deviate from the prescribed guideline of not providing model solutions, despite being instructed not to do that \cite{Hellas_2023}\cite{codehelp23}. Consequently, we implemented a self-assessment check wherein Iris verifies the adherence of its generated response to the predefined role of an "excellent tutor." If the response fails to conform to the rules, Iris refines the response or reduces the level of assistance until it aligns with the desired criteria.
\end{enumerate}

\section{Methodology}

This study revolves around the research questions (RQ1, RQ2, and RQ3) formulated in Section 1. We conducted an online survey among Iris' users to gather feedback on their experiences and perceptions regarding Iris' impact. In this study, we do not aim to measure the actual impact of Iris on student performance or learning behavior but rather focus on students' subjective perceptions and experiences. Follow-up studies will investigate the impact of Iris on student learning outcomes.

\subsection{Survey Design}

We asked the students to indicate their agreement with a series of statements on a five-point Likert scale ranging from "strongly agree" to "strongly disagree." These are the statements that relate to each research question:

\begin{enumerate}[label=\textbf{RQ\arabic*}, leftmargin=*]
  \item \textbf{Perceived Impact}
  \begin{enumerate}[label=\textbf{Q\arabic*}, leftmargin=*]
    \item Iris understands my queries well.
    \item Iris provides assistance that directly helps me with the issues I have while working on a programming exercise.
    \item The guidance offered by Iris has improved my understanding of programming concepts.
    \item Interacting with Iris makes the learning process more engaging.
    \item I feel more motivated to work on programming exercises when using Iris.
  \end{enumerate}
  
  \item \textbf{Preference of Iris Over Human Assistance}
  \begin{enumerate}[label=\textbf{Q\arabic*}, start=6, leftmargin=*]
    \item I feel comfortable asking Iris questions without worrying about being judged.
    \item I feel safe asking Iris questions that I wouldn't have the confidence to ask a tutor or professor.
    \item I prefer to ask questions to Iris instead of asking a human tutor for help.
    \item I would prefer to ask Iris questions about lecture content instead of asking the professor.
  \end{enumerate}
  
  \item \textbf{Reliance}
  \begin{enumerate}[label=\textbf{Q\arabic*}, start=10, leftmargin=*]
    \item I would find it more challenging to solve programming exercises without Iris.
    \item I find it difficult to solve the tasks in computer-based exams without Iris.
  \end{enumerate}
\end{enumerate}

The survey included additional questions specifically aimed at students who were aware of Iris but had yet to use it. These questions sought to understand the reasons behind their decision.

\subsection{Data Collection}
We surveyed students enrolled in three distinct CS1-level courses at the Technical University of Munich. These introductory courses aim to provide first-semester students of different study programs with programming fundamentals. Table \ref{tab:courses} provides an overview of the number of exercises with Iris enabled (\(n_{ex}\)), students enrolled (\(n_{st}\)), students who engaged with Iris at least ten times (\(n_{Iris}\)), conversations started with Iris (\(n_c\)) and messages sent to Iris in total (\(n_m\)).

\begin{table}[h]
  \caption{Overview of courses}
  \label{tab:courses}
  \begin{tabular}{lcrrrrrr}
    \toprule
    Study Program & $n_{ex}$ & $n_{st}$ & $n_{Iris}$ & $n_c$ &$n_m$ \\
    \midrule
    Management \& Tech. & 50 & 403 & 136 &  1629 & 7562 \\
    Informatics & 64 & 1141 & 72 &  1063 & 3430 \\
    Information Engineering & 10 & 111 & 13 &  109 & 408 \\
    \bottomrule
  \end{tabular}
\end{table}

We conducted the survey using LimeSurvey\footnote{\url{https://www.limesurvey.org}}, an open-source survey application. It was distributed to students via email and was open for a period of ten days.

\subsection{Data Analysis}

We chose a quantitative approach for the analysis of the survey data. The data cleansing process involved filtering responses to ensure completeness for questions Q1 to Q11. Additionally, participants were filtered based on the number of messages sent, with only responses from students who had sent a minimum of ten messages being included to ensure the evaluation of Iris' effectiveness is based on informed judgments. Finally, 26\% of the initial sample remained for further analysis.

221 students engaged with Iris by sending at least ten messages. Of these students, 121 successfully participated in the survey, resulting in a relative response rate of 55\%.
It is important to note that human tutors were available to all students in each course, which may have contributed to lower usage rates of Iris.
We employed a stacked bar chart as a visual representation to depict the distribution of responses for each question.

\begin{figure*}[t]
  \centering
  \includegraphics[width=1\textwidth]{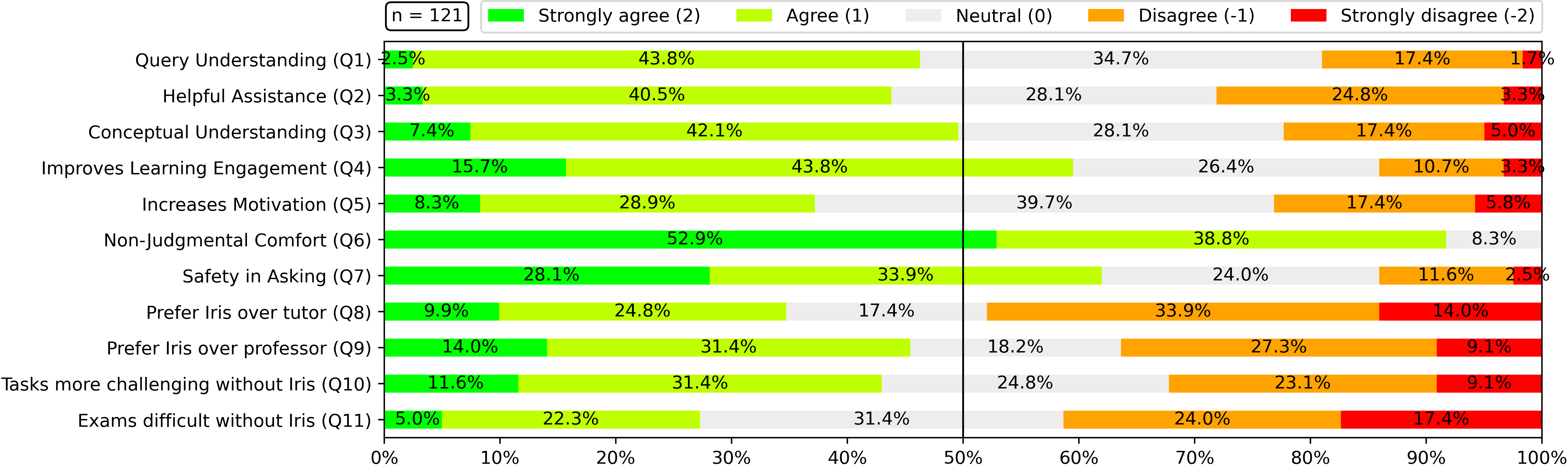}
  \vspace{-6.5mm}
  \caption{Stacked bar chart showing the distribution of responses for each question on a five-point Likert scale}
  \vspace{-4mm}
  \label{fig:results}
\end{figure*}

\section{Results}

In the following paragraphs, we present the results. The answers to each question are visualized in Figure \ref{fig:results}.

46\% of students reported that Iris comprehends their inquiries well, with 35\% neutral and 19\% disagreeing (Q1). For direct assistance, 44\% agreed, 28\% were neutral, and 28\% disagreed (Q2).  The enhancement of understanding in programming (Q3) received a positive response, with 50\% agreeing, 28\% neutral, and 22\% disagreeing.

Responses varied more on engagement and motivation. 60\% found interactions with Iris engaging, with 14\% disagreeing and 26\% neutral (Q4). Motivation responses were evenly spread, with 37\% agreeing, 40\% neutral, and 23\% disagreeing (Q5).

A significant 92\% felt comfortable asking Iris questions without judgment (Q6). 62\% felt safe asking Iris sensitive questions, whereas 14\% disagreed (Q7). For Q8, 35\% agreed, 48\% disagreed, and 17\% remained neutral. For Q9, 46\% agreed, while 36\% disagreed, and 18\% were neutral.

Regarding the reliance on Iris, 43\% believed it would be challenging to solve programming exercises without it, 32\% were neutral, and 24\% disagreed (Q10). For computer-based exams, 27\% thought tasks would be difficult without Iris, 31\% were neutral, and 41\% disagreed (Q11).

\section{Findings}

The analysis of responses to RQ1 suggests a generally positive perception of Iris' ability to understand student queries and provide assistance in resolving programming exercise issues. Most participants agreed that Iris understands their queries well and provides helpful assistance, indicating effectiveness in its contextual understanding and feedback mechanisms. The response also indicates that Iris has a notable impact on improving students' understanding of programming concepts. However, the relatively high percentage of neutral responses suggests that while students perceive Iris as helpful, there might be room for enhancing its capabilities to ensure a more universally positive reception.

Regarding the engagement and motivational aspects of Iris, the responses were generally positive, with a majority agreeing that Iris makes the learning process more engaging. However, the motivation to work on programming exercises with Iris received a more balanced response, suggesting that while Iris contributes positively to the learning experience, its influence on motivation varies among students.

\findingsbox{%
\textbf{Main Findings for RQ1}: Iris is perceived positively in understanding student queries and providing relevant assistance, contributing to an improved understanding of programming concepts. It enhances the learning experience by making it more engaging, although its impact on student motivation varies.
}%
\vspace{8pt}%

The responses to RQ2 reveal a strong level of comfort and safety in asking Iris questions, indicating a significant level of trust in the system, possibly due to the private and non-judgmental nature of the AI interaction. However, the responses are more balanced regarding preferring Iris over human support. While a notable percentage of students prefer Iris for its accessibility and immediate feedback, a more significant portion values interaction with human tutors. This data suggests that while Iris is a valuable tool for certain aspects of learning, students consider it a complement rather than a replacement for human tutors. On the contrary, the responses to Q9 indicate that students are more open to using Iris for lecture content questions over asking the professor during the lecture. This insight aligns with the comfort and safety aspect of Iris, as students may feel more comfortable asking Iris questions during the lecture than asking the professor directly in front of their peers.

\findingsbox{%
\textbf{Main Findings for RQ2}: Students express high comfort and safety in asking Iris questions. The preference for Iris over human tutors is balanced, highlighting its role as a complementary tool rather than a complete substitute for humans. Students are more open to using Iris as a replacement for asking questions in the lecture.
}%
\vspace{8pt}%

The responses to RQ3 show that a notable portion of students agreed that not having access to Iris would make solving programming exercises more challenging, indicating a certain level of reliance on Iris. However, a majority of students disagree or remain neutral about a higher difficulty of solving exam tasks without Iris. This suggests that while Iris is a welcome resource for routine exercise-solving, students still feel confident that they can independently solve programming assignments in exams. In the courses covered in this evaluation, the instructors made students aware ahead of time that Iris is not allowed during exams, which might have influenced the responses to Q11 compared to Q10.

\findingsbox{%
\textbf{Main Findings for RQ3}: There is a moderate level of reliance on Iris for routine programming exercises, but the reliance decreases in the context of exams. This indicates that students see Iris as a helpful tool for practice, learning, and homework. However, students appear confident in their abilities to perform in exams without Iris.
}%

\section{Discussion}

Integrating Iris into Artemis has provided insightful lessons on the implementation and impact of AI-driven virtual tutors in educational settings.
Using a prompt to define the "excellent tutor" role for Iris was a key element in shaping the chatbot's behavior. This approach ensured that the AI provided calibrated assistance through subtle hints and counter-questions, aligning with the educational goal of fostering independent problem-solving.

The mixed-positive responses in perceived impact suggest that the effectiveness of these prompts can be further optimized. Future iterations could benefit from refining the role definition and the Chain-of-Thought processes, including enhanced access to context. The current approach of presenting the model with a list of files to choose from is suboptimal, requiring the model to decide which files to look into based on the file name. It could be enhanced by building an embedding index of the files and their content and providing the model with the exact portions of the code and the problem statement relevant to the student's question. 

The reliance on Iris for routine tasks may have unintended consequences on learning habits and critical thinking skills. Over-reliance on AI assistance could lead to a lack of deep engagement with the material or diminished problem-solving skills, as students might opt for the path of least resistance rather than dealing with challenging concepts themselves. However, as the results suggest that students feel less reliant on Iris in exam contexts, the AI's role as a supplemental resource is well-established, and students know they need to be able to solve tasks without Iris. However, this perception may be influenced by the instructors informing students at the beginning of the course that Iris will not be allowed during exams. We recommend communicating the intended use and limitations of AI tooling to students in advance to ensure they understand its role in their learning process.

In the study, we selectively analyzed responses from students who had engaged with the system through a minimum of ten messages based on the premise that a certain interaction threshold is necessary for providing informed feedback. Although this criterion may introduce a bias towards users demonstrating higher engagement levels, it allowed us to gather insights from users who have meaningfully integrated Iris into their learning journey. The data reveals that the agreement rate on Q2 for students engaging less than ten times was 9 percentage points lower than that of their more engaged counterparts. This discrepancy shows the critical role of sustained interaction in fully realizing the tool's potential and underscores the need for further research into optimizing initial interactions with Iris.

While not a focus of this study, some students did not use Iris at all. In their feedback, they indicated that they preferred using other resources or tools for assistance or were already satisfied with their current methods of learning and problem-solving and did not require additional assistance. Artemis provides feedback through automated test results, which may already provide sufficient assistance for this group of students. Further research is needed to explore how Iris might still offer unique value or complement existing methods, even for students who perceive no current need for additional resources.

The selection of the language model used in Iris may have implications for the quality of support provided. This study employed GPT-3.5-Turbo as a cost-effective solution. However, GPT-3.5-Turbo lacks the advanced reasoning capabilities of more recent models such as GPT-4 or GPT-4-Turbo. While it is worth exploring the potential improvements, the ten times higher costs associated with GPT-4-Turbo pose a practical limitation for large-scale educational settings. The cost of providing Iris during the winter semester of 2023/2024 already amounted to a substantial sum of 1500 euros for about 11,000 interactions. 

While students tend to appreciate the assistance from Iris, the varying degrees of reliance and the preference for human interaction in specific contexts suggest that the highest quality AI may only sometimes be necessary. An optimal solution might involve a hybrid model, where cost-effective AI solutions handle routine queries, supplemented by higher-quality AI or human intervention for complex or sensitive issues, ensuring efficiency and privacy.

It is important to navigate ethical concerns, such as ensuring data privacy and promoting equitable access, to maintain trust and fairness in each student's educational journey. Iris is free of charge for students. This promotes inclusivity and eliminates any potential barriers that may hinder students from seeking assistance from AI tooling. This democratization of access to educational support contributes to a more equitable learning environment where every student has an equal opportunity to excel in their programming exercises. 

\section{Limitations}

In line with Runeson and Höst's categorization framework \cite{Runeson2009guidelines}, we recognize potential limitations impacting the internal, external, and construct validity of this study:

\textit{Internal Validity}: 
Self-reported survey data may introduce biases, with perceptions potentially influenced by individual attitudes or varying familiarity with programming concepts. Notably, perceived effectiveness does not necessarily equate to objective effectiveness.
The result analysis focused on students who interacted with Iris at least ten times. This approach may overlook the perspectives of students who used the system less frequently, introducing a potential selection bias.

\textit{External Validity}: 
The findings, rooted in a specific educational setting, may not be broadly applicable due to unique factors like student demographics, course structure, and institutional policies.

\textit{Construct Validity}: 
The survey questions meant to evaluate 'perceived effectiveness' and 'comfort with asking questions' may not capture the full range of what they aim to assess. Underlying factors such as prior experiences or personal preferences, which are not captured by the survey, might influence these perceptions.

\section{Conclusion}

Students perceive Iris' personalized and context-aware assistance positively. Iris tends to aid in understanding programming concepts and engaging learners, although its impact on the motivation for programming exercises varies individually. While valued for practice, students indicate that they confidently rely on their skills for exams, suggesting a judicious use of the tool.
Most surveyed students trust and feel comfortable using Iris for queries but do not view it as a complete replacement for human interaction. While they value interactions with human tutors, they are more inclined to discreetly ask questions using Iris during lectures.

Future research should deepen the evaluation of Iris' context awareness by contrasting Iris with general-purpose AI tools like ChatGPT in a controlled experimental setting.
This experiment should involve three distinct groups of students: a control group not using any AI tools, a group using Iris, and a group using a general-purpose AI tool. Pre- and post-tests should be conducted to quantitatively assess the impact on learning outcomes. Further analysis should assess the number of rejected questions and the number of actually valuable responses, providing insights into the question filtering mechanism's effectiveness and Iris' response quality.
Moreover, future studies should explore the factors that deter student engagement with Iris and develop further strategies to enhance the learning experience with Iris.

Exploring the integration of different LLMs like GPT-4-Turbo, a fine-tuned GPT-3.5-Turbo, or open source models such as Llama2 could provide insights into the trade-offs between quality, data privacy, and cost.
Enhancing Iris' integration with student code is crucial as well. Rather than selecting files by name, developing an embedding index for file content will enable Iris to access and use precise code segments and related parts of the problem statements.

\bibliographystyle{ACM-Reference-Format}
\balance
\bibliography{base}

\end{document}